\documentclass[letterpaper, 11 pt, conference]{IEEEtran}

\usepackage{graphicx} % for pdf, bitmapped graphics files
\usepackage{bm}\usepackage{graphicx}
\usepackage{hyperref}

\title{\LARGE \bf Blockchain-based certificate authentication system with enabling correction}

\makeatletter
\newcommand{\linebreakand}{%
  \end{@IEEEauthorhalign}
  \hfill\mbox{}\par
  \mbox{}\hfill\begin{@IEEEauthorhalign}
}
\makeatother 

\author{
  \IEEEauthorblockN{Md. Mijanur Rahman}
  \IEEEauthorblockA{\textit{Dept. of Computer Science and Engineering} \\
    \textit{Southeast University}\\
    \href{mailto:mijanur.rahman@seu.edu.bd}{mijanur.rahman@seu.edu.bd}}
  \and
  \IEEEauthorblockN{Saifur Rahman Shihab}
  \IEEEauthorblockA{\textit{Dept. of Computer Science and Engineering} \\
    \textit{Southeast University}\\
    \href{mailto:rahmansaifur223@gmail.com}{rahmansaifur223@gmail.com}}
  \linebreakand % <------------- \and with a line-break
  \IEEEauthorblockN{Md Tanzinul Kabir Tonmoy}
  \IEEEauthorblockA{\textit{Dept. of Computer Science and Engineering} \\
    \textit{Southeast University}\\
    \href{mailto:tanzinulkabir@gmail.com}{tanzinulkabir@gmail.com}}
  \and
  \IEEEauthorblockN{Riya Farhana}
  \IEEEauthorblockA{\textit{Dept. of Computer Science and Engineering} \\
    \textit{Southeast University}\\
    \href{mailto:riyafarhana56@gmail.com }{riyafarhana56@gmail.com }}
}

\begin{document}

\maketitle

%%%%%%%%%%%%%%%%%%%%%%%%%%%%%%%%%%% Abstract %%%%%%%%%%%%%%%%%%%%%%%%%%%%%%%%%%%%%%%%%%%%%

\textbf{\textit{Abstract}--Blockchain has proven to be an emerging technology in the digital world, changing the way everyone thinks about data security and bringing efficiency to several industries. It has already been applied to a wide range of applications, from financial services and supply chain management to voting systems and identity verification. An organization must verify its candidates before selecting them. Choosing an unqualified candidate can ruin an organization's reputation. In this digital era, many key fraudulent schemes are rampant in many companies and one of them is certificate fraud. It is possible to validate a candidate's qualifications using traditional methods, but there are drawbacks such as security issues and time consumption. In this paper, a blockchain-based academic certificate authentication system will be used to ensure authenticity and make the assertion of the decentralized system secure. However, the system will generate, authenticate and make corrections on academic certificates. Ultimately, some blockchain-based authentication systems already exist, they can't correct any errors that occur during generation. The proposed system will help in many ways, such as providing a user-friendly university admission, and smooth job hiring process, etc. In conclusion, our proposed system can permanently eradicate certificate forgeries and create and promote trust in society.} \\


%%%%%%%%%%%%%%%%%%%%%%%%%%%%%% Introduction %%%%%%%%%%%%%%%%%%%%%%%%%%%%%%%%%%%%%%%%%%%%%%%%%%
\section{{\Large I}NTRODUCTION}

In our daily lives, certificates play an invaluable role [1]. It is a document that is provided to a student when he/she graduates from a school, college or university. Human capital is defined as the knowledge, skills, and abilities gained via education [2]. A student's credentials serve as a testament to his or her diligent work over the years. Those graduate students need these certificates to apply for higher education or to obtain employment in respectable places. In both circumstances, they need to provide their credentials for validation. To validate, institutes check the authenticity of certificates provided by candidates. After validation, candidates proceed to the next stage. To validate the authenticity of a certificate, the institute uses a third-party system or direct contacts with the institute from which it was generated which is often a time-consuming and complicated process. \\

There are currently 2 million fake degree certificates in circulation in the United States and 300 unauthorized universities operating [3]. If proper validation was not conducted, then any unworthy candidates who had resorted to forgery of their qualifications would be hired. The reason behind certificate forgery is that many unethical people want to get their dream job without having the necessary qualifications, so they provide fake certificates in order to deceive the hiring company. This makes the validation process much more difficult and time-consuming, as the hiring company must now carefully scrutinize the qualifications of each candidate to ensure their legitimacy. A study shows that almost 10$\%$ of applications given by candidates are forged [4]. Applicants tend to lie about their education and experience [5]. Every year, academic certificate fraud costs employers approximately $\$$600 billion [6].\\

“The phenomenon of fake academic degrees is a threat to any community. It is an enormous threat, both for the present and the future” [7]. Nowadays there are numerous cases of certificate forging in the news. In recent years, certificate forgeries have increased due to advances in scanning and printing technologies. As a result, the integrity of both the certificate holder and the institute that issued the certificate is at risk. [8]. For 28 years, the Massachusetts Institute of Technology's dean of admissions was found to have faked and misled the institute about her academic qualifications [9]. \\

Certificate forging is also causing problems in the medical field. Many people pretend to be a doctor by forging fake certificates. As a result, people from all socioeconomic backgrounds will not receive equal treatment when they visit different types of doctors [10]. A similar report was published in Bangladesh. It was for using expired reagents and selling unapproved drugs. Seven employees, including two fake doctors, were sentenced to indeterminate jail terms and penalties for taking high charges from patients and providing fake medical reports [11]. In Mexico, there are some aesthetic pseudo-clinics with non-medical personnel performing surgery on patients which caused many complications [12]. \\

Blockchain is one of the most popular technologies that transform the way we live at the moment. Blockchain is a decentralized database that contains records called a block. Each block contains its timestamp, hash or address of the previous block and the data of the block. Each block has a unique hash that can be used to track back to its previous block. A hash is a function that converts a set of data into a fixed-size data structure, called a hash value. Blockchain technology has several applications, including cryptocurrency, smart contracts, supply chain management, and more. Stuart Haber and W. Scott Stornetta, two mathematicians who wanted to implement a system where document timestamps could not be tampered with, proposed blockchain technology in 1991 [13]. Since the information on the blockchain is unchangeable, it gives assurance that no one can tamper with the information. If any attempt is made, it can easily be identified because of the changes in the hash value [13]. As the authors point out [14], traditional authentication methods do not provide security, tamper-proofing, or authentication for documents. For this reason, blockchain is best suited to our system. \\

Based on this study, we proposed a system to resolve a crucial pair of problems in the world of education and job hiring using blockchain technology. The system will verify whether the certificates are given by any legitimate organization or academy. The system will permanently store all the credentials securely, making the authentication process much easier and more convenient, and eliminating the crime of certificate forging. The system will also have correction functionality so that if any modifications need to be done, the authority can easily accomplish them. \\

%%%%%%%%%%%%%%%%%%%%%%%%%%%%%% Introduction %%%%%%%%%%%%%%%%%%%%%%%%%%%%%%%%%%%%%%%%%%%%%%%%%%

%%%%%%%%%%%%%%%%%%%%%%%%%%%%%% LITERATURE REVIEW %%%%%%%%%%%%%%%%%%%%%%%%%%%%%%%%%%%%%%%%%%%%%%%%%%
\section{{\Large L}ITERATURE REVIEW}

Blockchain was introduced long before Bitcoin was introduced by Satoshi Nakamoto[15], [16]. However, blockchain gained its popularity with Bitcoin. Although bitcoin is sometimes referred to as a blockchain by some people, but the fact is bitcoin was built using blockchain technology. In a short period of time, blockchain became popular when people discovered its salient benefits. It has become one of the top technologies worldwide. For this reason, many studies and research were conducted and this proposed system rewired some of them. \\

Authors of [17] have made some excellent points about blockchain architecture and consensus algorithms. They have also examined the limitations of blockchain and given some possible solutions. They also investigate various consensus algorithms in different respects and discuss their differences. By analyzing those differences, they can determine the most appropriate consensus algorithm. \\

The author of [4] Rishabh Garg, has examined how blockchain technology can be applied to solve problems within academic institutes and the employment sector. According to the author, blockchain will be the panacea for this issue. He proposed a framework that can issue new certificates as well as validate them. \\

In Osman Ghazali and Omar S. Sale [18], emphasized the significance of certificate verification and its influence in our society. In the paper, traditional verification processes were briefly explained, as well as their limitations. To eliminate those limitations, they proposed a blockchain-based verification system. A system that is not only used for verification but also for generating new certificates. They will generate digitally signed certificates using the asymmetric key and timestamp. Students will receive a copy, which they can verify in the system. \\

Jayesh G. Dongre and his colleagues proposed[19] to solve the problems of the current system of certificate verification. In the paper, they talked about the current verification process and the proliferation of certificate fraud. Using blockchain, they have developed a platform for validating and generating certificates. In their view, the use of blockchain for certificate verification is beneficial to society. It will eliminate certificate fraud. \\

R.Suganthalakshmi[20] and her colleagues investigated an optimal method of verifying academic credentials in 2022. The number of graduating students is increasing every year, and it is becoming increasingly necessary to validate their academic credentials. Without proper validation, any ineligible candidate will be blessed with opportunities. Their solution is to create a platform for all certificates a student may possess. Students upload all their credentials to the system, and the system stores them on the blockchain. To verify a credential, a person needs the student's ID and password. They used the PoW(Proof of work) consensus algorithm for validation, but the validation method was not well elucidated. \\

In the papers mentioned, researchers are interested in building a system to verify and generate academic credentials. Although their system design is different from each other but they all agree on using blockchain for security. Our proposed system can also generate and verify academic credentials, as well as modify them. Making errors is a part of human nature. Our study found that none of the papers offered editing options if any modifications were needed. We provide an editing facility option to fix those errors.

%%%%%%%%%%%%%%%%%%%%%%%%%%%%%% LITERATURE REVIEW %%%%%%%%%%%%%%%%%%%%%%%%%%%%%%%%%%%%%%%%%%%%%%%%%%

%%%%%%%%%%%%%%%%%%%%%%%%%%%%%% METHODOLOGY %%%%%%%%%%%%%%%%%%%%%%%%%%%%%%%%%%%%%%%%%%%%%%%%%%
\section{{\Large M}ETHODOLOGY}

In this study, the general idea of the proposed system is depicted in Fig-1. There will be two actors, the university (Admin), and other users (Students/Employer). The university is authorized to generate new certificates for students, make corrections if necessary, and authenticate the certificates. On the other hand, general users are allowed only to confirm the authenticity or view the certificates. The user in this category can't make any corrections or generate new certificates. We utilize two chains in our system. One is for storing certificate data and the other is for keeping track of corrections.
% \begin{figure}
%     \centering
%     \includegraphics{A.png}
%     \caption{Architecture}
%     \label{fig:Architecture}
% \end{figure}
\begin{center}
    \includegraphics[width=\linewidth]{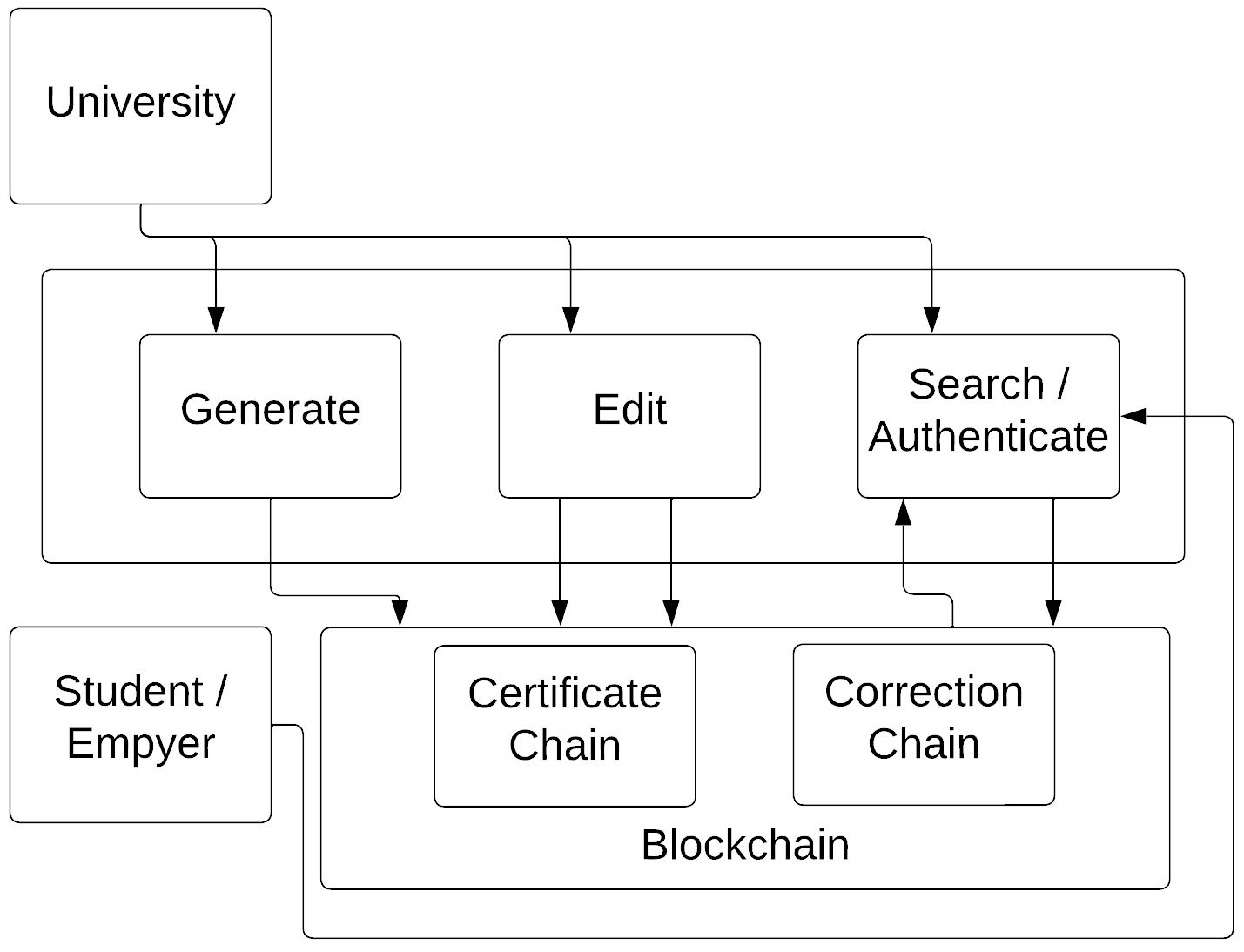} \\
    Figure 1: System Architecture
\end{center}

% \begin{figure}[h]
%   \includegraphics[width=\linewidth]{A.png}
%   \caption{System Architecture.}
%   \label{fig:Architecture}
% \end{figure}

% \noindent \\ 
The certificate generation process is shown in Fig-2. In this system, everyone can check the authenticity of the certificates but not everyone can generate or add new certificates to the system. Only authorized persons will be allowed. First, an authorized person needs to prove their identity by login in. After proving their identity, they can add a new certificate to the blockchain. To add a certificate, the user may need to fill up a form. In that form, they have to provide the information of the candidate. After hitting submission, the system will generate a new block and store the data on the certificate chain. After that, it will return the block address to the university and the student.
\begin{center}
    \includegraphics[width=\linewidth]{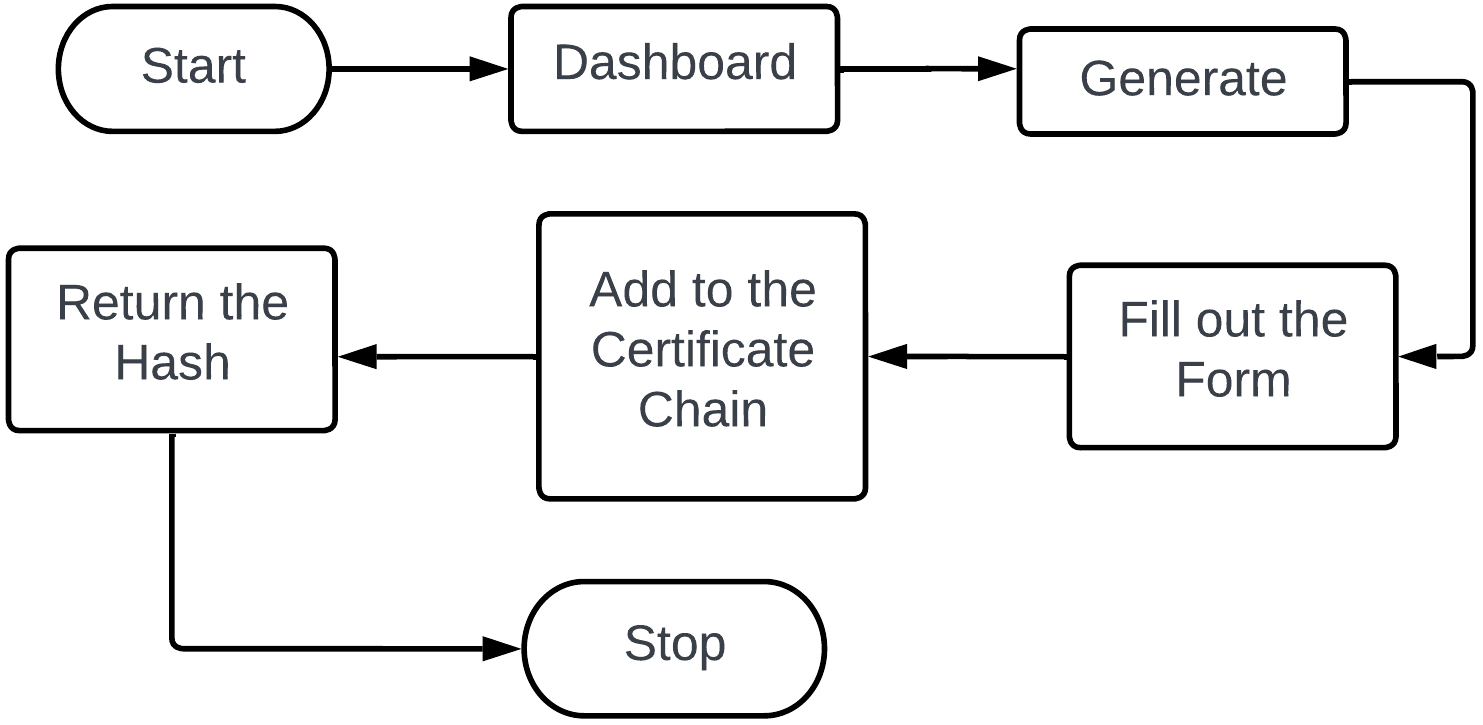}
    Figure 2: Generating new certificates.
\end{center}
% \begin{figure}[h]
%   \includegraphics[width=\linewidth]{G.png}
%   \caption{Generating new certificates.}
%   \label{fig:Generating}
% \end{figure}

\begin{center}
    \textbf{PESUDO CODE 1:} \textit{Function to generate Certificate}
\end{center}

{\noindent function generate Certificate(Arguments) public \{}
 \\ {\indent check if authorized to generate a Certificate}
 {\indent then \\ \indent store the certificate 
return the address of} the certificate \\
\} \\

The certificate correction process is depicted in Fig-3. To correct a certificate, we need to provide the system with the address of the old block and the updated data. After getting those two pieces of information our system will generate two new blocks.  The first block will contain corrected data and it will be stored in the certificate chain.  The second block will contain the hash address of the old certificate and the hash address of the corrected block. However, the modified block will be stored in the correction chain as a reference point.
\begin{center}
    \includegraphics[width=\linewidth]{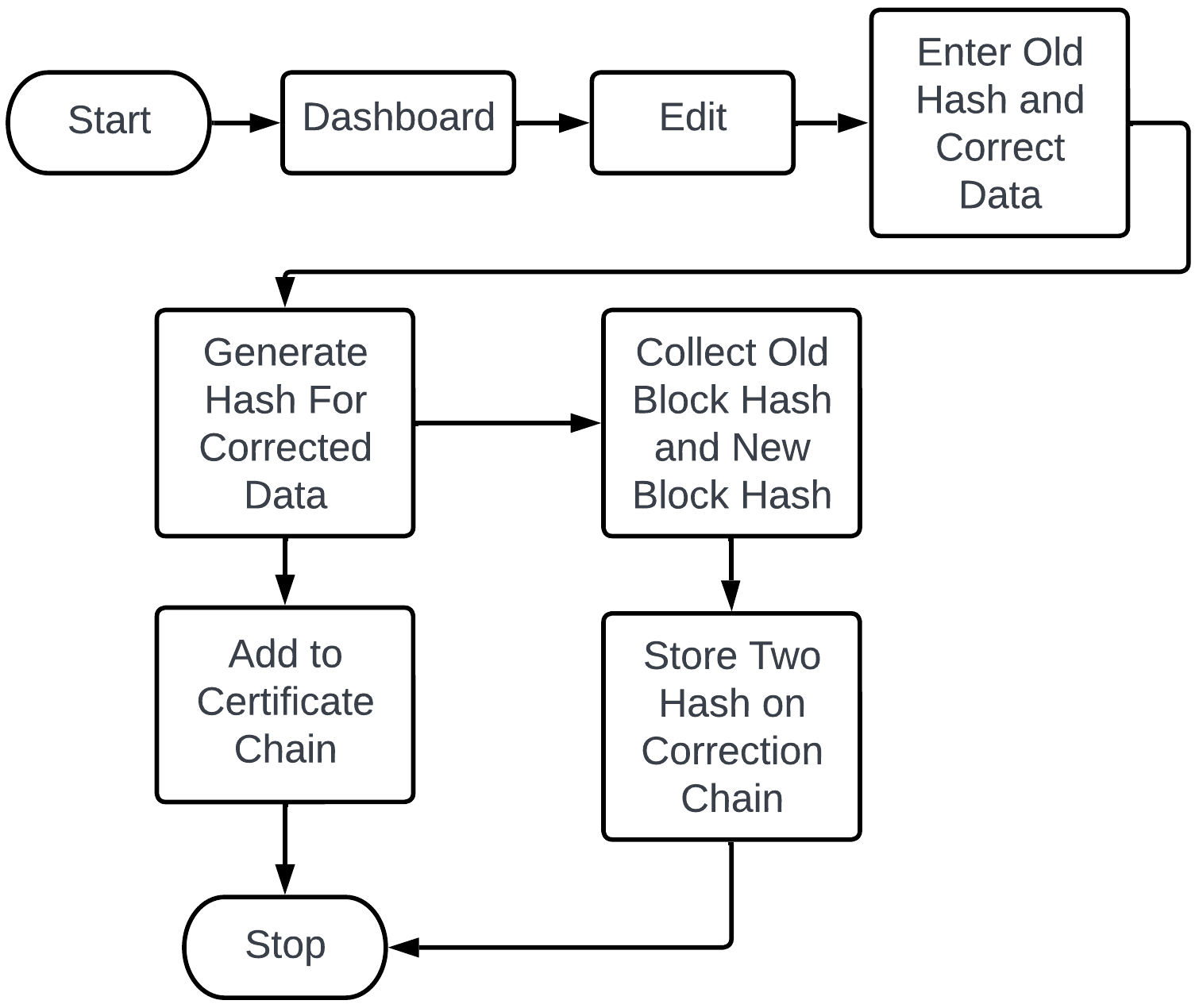}
    Fig-3: Certificate correction
\end{center}
\begin{center}
    \textbf{PESUDO CODE 2:} \textit{Function to make correction}
\end{center}

function correct Certificate (Argument old\_Certificate\_addressother arguments) public \{ \\
{\indent Check if authorized  to perform the correction.} \\ {\indent \indent Then check if old\_Certificate\_address exists} {\indent \indent on the certificate chain.Store new certificate on} {\indent certificate chain. \\ \indent \indent Store old and new certificate} {\indent addresses on the correction chain.\\ \indent return the} {\indent corrected certificate address.} \\
\} \\

The validation-checking process is shown in Fig-4. In our system, a certificate can be checked in two ways. This can either be done by manual typing or scanning the QR code. QR code will contain the hash address of the block where the data is stored. Apparently, the system will look for the hash address in the correction chain. Upon finding a match, the system will get the corresponding new certificate hash from the block. Later, it will search for the hash in the certificate chain and display the result. When the system is not able to locate the hash address in the correction chain, it searches the certificate chain and displays the result. If the system could not locate the hash address in both chains, then it will show certificate does not exist.
\begin{center}
    \includegraphics[width=\linewidth]{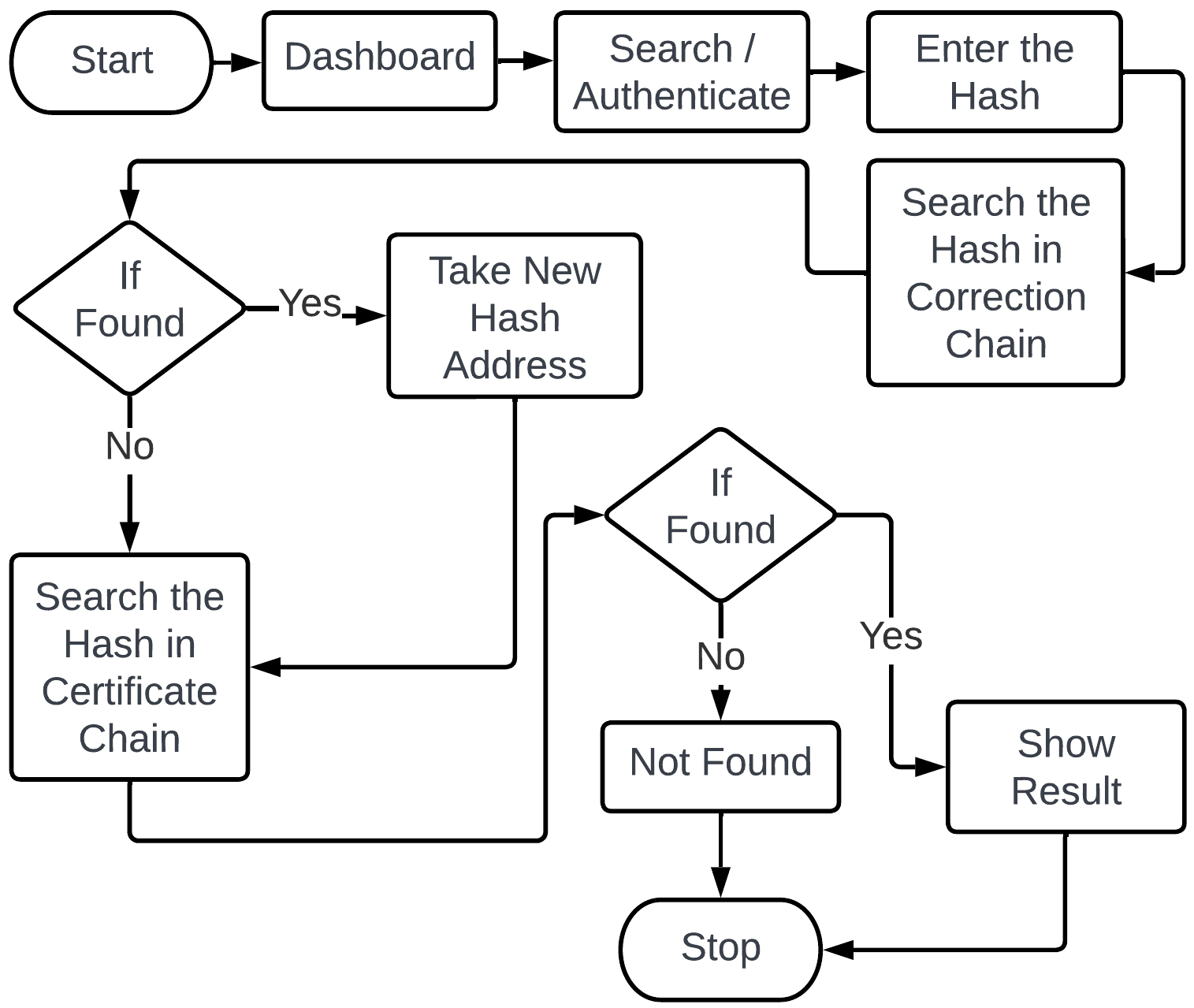}
    Fig-4: Checking the authenticity
\end{center}
\begin{center}
    \textbf{PESUDO CODE 3:} \textit{Function to Authenticate}
\end{center}

function authenticate (Argument address) public \{ \\
{\indent check if exists on the correction chain}  \\ 
{\indent \indent then} \\
{\indent \indent \indent get a new certificate address and check if \\ \indent \indent \indent a new certificate address exists on the \\ \indent \indent \indent certificate chain} \\
{\indent \indent then return the corrected certificate.} \\
{\indent \indent \indent else return "Not Exist"} \\
{\indent else check if exists on the certificate chain}
then \\
return the certificate \\
else return "Certificate does not exist" \\
\} \\

\textbf{SHA-256:} SHA-256 is a part of SHA-2 (Secure Hash Algorithm 2). It is a popular hashing algorithm. A cryptographic hash, known as a fingerprint, or signature, is a nearly identical string of characters generated from a different piece of input text. SHA-256 generates a 256-bit signature. SHA-2 is widely known for its security (it hasn't weakened as much as SHA-1) [21]. \\

\textbf{Smart Contract:} Nick Szabo introduced this concept in 1994 and defined it as a smart contract [22]. A smart contract is nothing but a computer procedure that verifies, facilitates, or enforces renegotiation or efficiency. Smart contracts enable the execution of credible exchanges without the involvement of outsiders such as lawyers or notaries. On the blockchain, smart contracts execute exactly as programmed, with no potential for censorship, shutdown, forgery, or third-party interface [23].
Smart contracts operate by executing basic "if/when...then..." phrases encoded into blockchain code. When preset conditions are met and validated, a network of computers executes the actions. When the transaction is finished, the blockchain will be updated. This indicates that the transaction is final and that only parties to whom permission has been granted can view the outcome [24]. \\

\textbf{Solidity:} Solidity is a high-level contract-oriented language for creating a smart contract. It was inspired by C++, Python, and JavaScript and is intended for use with the Ethereum Virtual Machine. Solidity is a programming language that is similar to OOP (object-oriented programming) [12]. The significance of Solidity programming for the Ethereum blockchain is based on the ability to develop industry-grade blockchain applications. The Ethereum Network team created it specifically for designing and creating smart contracts on blockchain platforms [21].

%%%%%%%%%%%%%%%%%%%%%%%%%%%%%% METHODOLOGY %%%%%%%%%%%%%%%%%%%%%%%%%%%%%%%%%%%%%%%%%%%%%%%%%%

%%%%%%%%%%%%%%%%%%%%%%%%%%%%%% RESULT AND DISCUSSION %%%%%%%%%%%%%%%%%%%%%%%%%%%%%%%%%%%%%%%%%%%%%%%%%%
\section{{\Large R}ESULT AND {\Large D}ISCUSSION}
In our proposed system, we adopted Remix IDE for writing and developing our smart contract. For testing, we deployed the smart contract on Remix's virtual network. Our study estimated three primary functions: (1) Generate Certificate, (2) Correct Certificate and (3) Authenticate Certificate. Upon deploying the smart contract in the Remix virtual machine, it gives us information about the transaction hash, gas fee, and more which is illustrated in Fig-5.
\begin{center}
    \includegraphics[width=\linewidth]{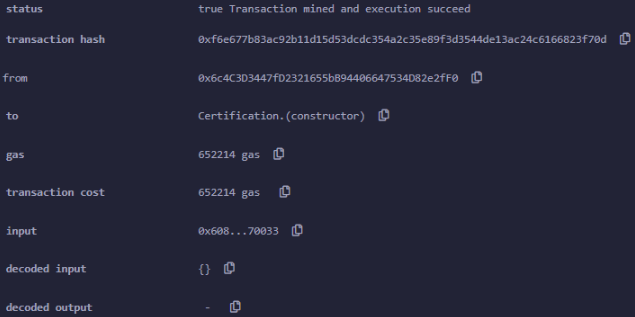}
    Fig-5: Deploy on Remix IDE
\end{center}

After successfully testing the smart contract on Remix's virtual network, we deploy the code on the Goerlitestnet network. During the deployment phase, it asks for confirmation. Our metamask wallet confirms it. After successfully deploying the smart contract on the blockchain, we obtained the address and the ABI (Application Binary Interface) of the smart contract from Remix IDE and the medium of our front-end connection with the blockchain. 
\begin{center}
    \includegraphics[width=\linewidth]{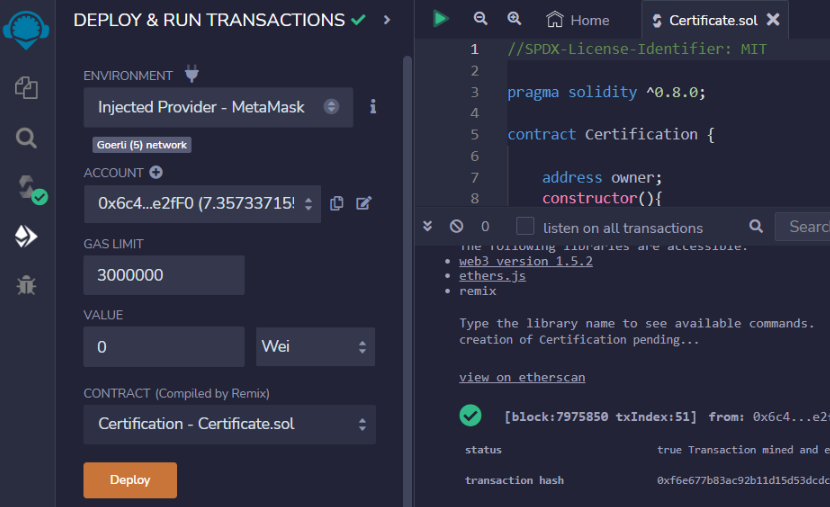}
    Fig-6: Deploy to Goerlitestnet
\end{center}

\begin{center}
    \includegraphics[width=\linewidth]{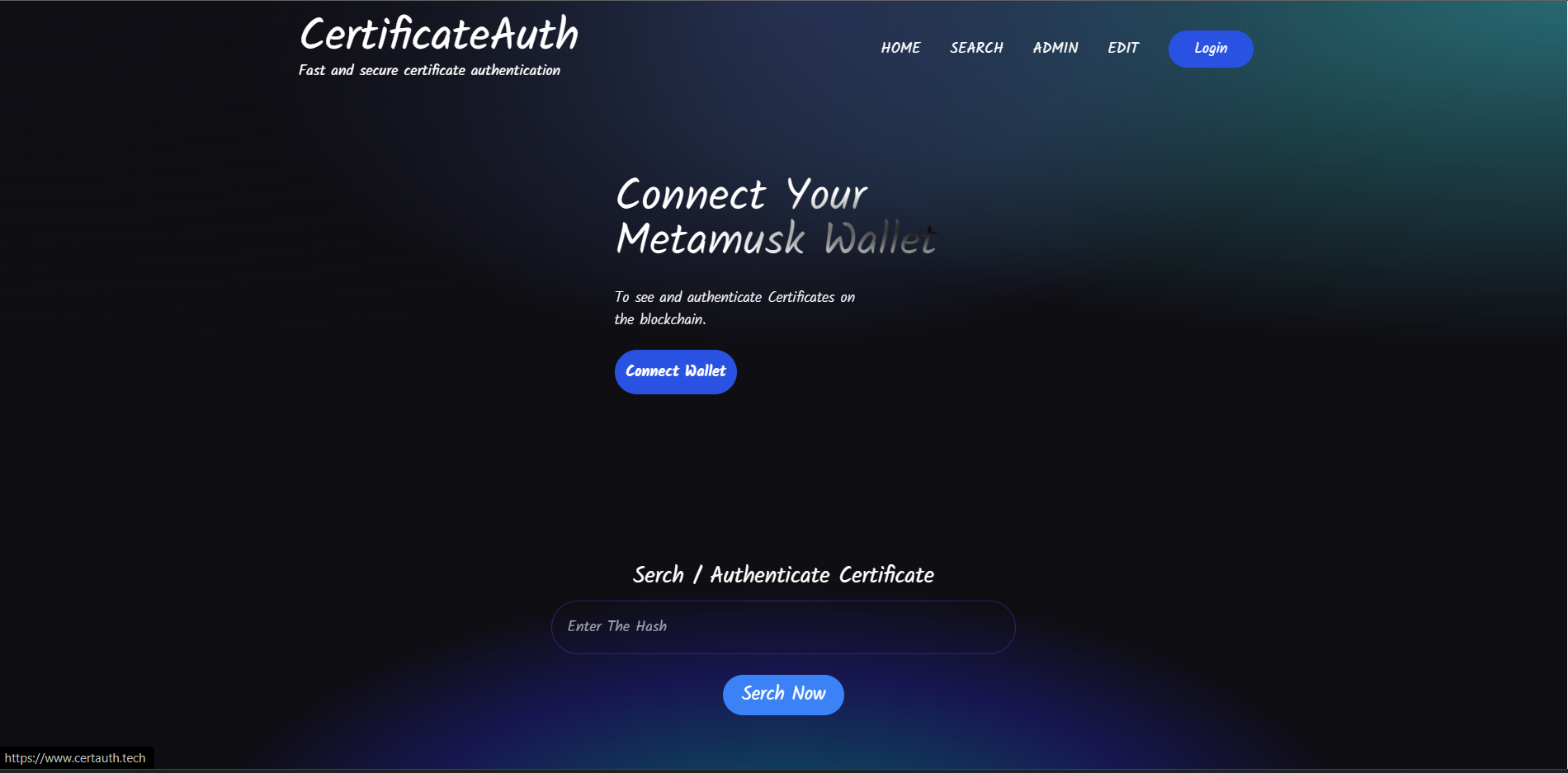}
    Fig-7: Web Application
\end{center}
Fig-7 shows the frontend web application of the proposed system.
% \begin{figure}[h]
%   \includegraphics[width=\linewidth]{logo.png}
%   \caption{Web Application.}
%   \label{fig:Application}
% \end{figure}
%%%%%%%%%%%%%%%%%%%%%%%%%%%%%% RESULT AND DISCUSSION %%%%%%%%%%%%%%%%%%%%%%%%%%%%%%%%%%%%%%%%%%%%%%%%%%

%%%%%%%%%%%%%%%%%%%%%%%%%%%%%% CONCLUSION %%%%%%%%%%%%%%%%%%%%%%%%%%%%%%%%%%%%%%%%%%%%%%%%%%
\section{{\Large C}ONCLUSION}
Because many candidates fabricate certificates, traditional validation methods are facing new challenges in the digital age. Fabrication of certificates occurs in order to gain admission to prestigious universities or to obtain a satisfactory job. To overcome those challenges the validation process needs to be more advanced and secure. Blockchain technology is ideal for tackling those challenges. Due to the enormous characteristics of blockchain, such as decentralization, census, and security, it has become one of the leading technologies in recent years. Our proposed system is built on a blockchain architecture. It will not only validate certificates but also has the ability to generate brand-new certificates for students. Other blockchain-based verification platforms only offer verification and the ability to generate new certificates, but don't allow the correction of certificates. However, during the process of creating the certificate, a mistake can occur. If there is no way to make up for those mistakes, then it is not an ideal platform. In this proposed system if any mistakes were made, they can be fixed and by eliminating certificate forgeries and creating a trustworthy environment, the system will guarantee the integrity of certificates. \\

As the system develops, it can be expanded to include blockchain-based professional degree management, implementation of a NID (National Identity Card) verification system, and other blockchain-based systems.
%%%%%%%%%%%%%%%%%%%%%%%%%%%%%% CONCLUSION %%%%%%%%%%%%%%%%%%%%%%%%%%%%%%%%%%%%%%%%%%%%%%%%%%

%%%%%%%%%%%%%%%%%%%%%%%%%%%%%%%%%%% Reference  %%%%%%%%%%%%%%%%%%%%%%%%%%%%%%%%%%%%%%%%%%%%%

\end{document}